\documentclass[aip,jcp,reprint]{revtex4-1}
\usepackage[applemac]{inputenc}
\usepackage{amssymb}
\usepackage{times}
\usepackage{graphicx}
\usepackage{subfigure}
\usepackage{natbib}
\usepackage{mciteplus}
\usepackage{dcolumn}
\usepackage{bm}
\usepackage{mathrsfs}
\usepackage{multirow}
\usepackage{amsmath,amsthm,amsfonts,amssymb}
\usepackage{lipsum}
\usepackage{mathrsfs}
\usepackage[bottom]{footmisc} 
\usepackage{mathrsfs}
\usepackage{float}
\usepackage{xcolor}
\usepackage{soul} 

\begin{document}

\title{Amplification of the diamagnetic response in small Hubbard rings}

\author{T. V. Trevisan}
\affiliation{Instituto de F{\'i}sica Gleb Wataghin, Universidade Estadual de Campinas, UNICAMP, 13083-859 Campinas, S{\~a}o Paulo Brazil}
\author{A. O. Caldeira}
\affiliation{Instituto de F{\'i}sica Gleb Wataghin, Universidade Estadual de Campinas, UNICAMP, 13083-859 Campinas, S{\~a}o Paulo Brazil}

\begin{abstract}
In this paper, we present a study of the electric transport properties of small discrete rings with $3\leq N\leq 6$ sites and $N_{e}<2N$ electrons, which can be seen as a simplified version of real aromatic molecules. In particular, the ring with six sites and six electrons is our prototype of the benzene molecule. It is already known that the Hubbard model itself cannot account for the anisotropy of the diamagnetic susceptibility of the aromatic molecules, which is observed when they are subjected to an external magnetic field perpendicular to their basal plane. Therefore, we propose an extension of the Hubbard model, with an \textit{ad hoc} extra interaction term, with two adjustable parameters. Our results show that this extension of the Hubbard model is able to amplify the persistent currents established in the ground state of our rings and, moreover, promotes an enhancement of the magnetic susceptibility anisotropy depending on the tuning of the adjustable parameters. In particular, for the prototype of the benzene molecule, we recover the order of magnitude of the diamagnetic anisotropy measured for this molecule.  
\end{abstract}

\date{\today}

\maketitle

\section{Introduction}

More than $80$ years ago a peculiar property of the aromatic molecules was observed\cite{KrishnanI, KrishnanII}: these molecules exhibit, in the presence of a magnetic field perpendicular to their basal plane, a large diamagnetic anisotropy. In other words, the component of the molecule's magnetic susceptibility parallel to the magnetic field ($\chi_{\parallel}$) is much more intense (in absolute values) than the perpendicular components ($\chi_{\perp}$). The early efforts to explain this anisotropy were made by Linus Pauling \cite{Pauling}, Kathleen Londsdale \cite{Londsdale} and Fritz London \cite{LondonI,LondonII,LondonIII}, whose ideas constitute the essential features of the \textit{Ring Current Model}\cite{Lazzeretti} (RCM). 

Pauling argued that the $\pi$-electrons, which occupy the $p_{z}$ orbitals of the carbon atoms of the aromatic ring, are those that contribute more significantly to the susceptibility anisotropy in aromatic molecules. The density of probability to find these electrons is meaningful only in two ring-shaped regions: one above and other below the carbon ring of the aromatic molecule. In these regions, the $\pi$-electrons feel an interaction potential (due to the others electrons of the system and also the nuclei of the atoms) which is approximately cylindrically symmetrical with respect to the axis perpendicular to the basal plane of the molecule passing through its center (say, $z$ axis, for simplicity). In these circumstances, each of the $\pi$-electrons, in the presence of an external magnetic field parallel to the $z$ axis, contributes to $\chi_{\parallel}$ through the Pauli mechanism (Pauli susceptibility), which is proportional to the mean square distance of the electrons to the $z$ axis. With this model, Pauling estimated\cite{Pauling} the susceptibility anisotropy for the benzene molecule as being $\Delta\chi=-4,92\times 10^{-5}cm^{3}/mol$, not so far from the experimental value $\Delta\chi=-6,49\times 10^{-5}cm^{3}/mol$ \cite{And}. 

London, in turn, interpreted Pauling results in terms of a non-dissipative electric current along the aromatic ring, which he called a ``\textit{supracurrent}''\cite{LondonIII}. Because of the non-dissipative character of this ``\textit{supracurrent}'', it did not take long until researchers tried to trace a parallel between the ring currents in aromatic molecules and the supercurrent in a superconductor loop, expecting to find a microscopic model for the ring currents in aromatic molecules based on the BCS model \cite{Haddon}. However, it was not known at that time that mesoscopic normal metal rings could also carry an electric current without dissipation when subjected to an external magnetic field, provided that they were very pure and kept at very low temperatures. It is believed that the ring currents in aromatic molecules are of the same nature as these persistent current in mesoscopic normal metal rings \cite{Bouchiat,Hirsch} and they differ from the supercurrent in a superconductor loop in many aspects: the persistent current in normal metal rings is a genuine quantum effect due to the coherence of single electrons in the ring \cite{Imry} and, contrary to a supercurrent in a superconductor loop, they vanish if the external field is turned off. Besides, for rings with the same dimensions, the persistent current is much smaller than the supercurrent, since the latter is due to the collective motion of Cooper pairs condensed in a macroscopically occupied single quantum state. On top of that, whereas the supercurrent can only be diamagnetic, the persistent currents can be either diamagnetic or paramagnetic, depending on the number of electrons in the system\cite{Bouchiat}. Hereafter, every time we say \textit{persistent current}, we mean the non-dissipative electric current in small normal metal rings, instead of the supercurrents in superconductors.  

Until today the RCM model is broadly used as a criterion for aromaticity \cite{Aihara,Haddon} and it has been the subject of much controversy, as pointed out in the review article \cite{Lazzeretti}. 

Thus, motivated by the longstanding discussion concerning the ring currents in aromatic molecules, we investigate in this paper the electric transport properties of small discrete rings with $3\leq N\leq 6$ sites and $N_{e}\leq 2N$ electrons, which can be seen as a simplified version of real molecules (the $N$ sites of the rings represent the nuclei plus the core electrons of the atoms of the molecule while the $N_{e}$ electrons are the conduction $\pi$-electrons), being the ring with six sites and six electrons our prototype of the benzene molecule. We are interested in investigating whether the ground state of these rings can be a current-carrying state in the absence of any external field, and also quantify the persistent current that is established in the ground state of these systems under the influence of a uniform and static magnetic field perpendicular to the plane of the rings and in the Aharonov-Bohm configuration. Since we are interested only in the electronic degrees of freedom, the sites of the rings are always static. Besides, we consider only one non-degenerate electronic orbital per site and a more sophisticated multi-orbital theory will be explored in a future work. In section \ref{MeR}, after having analyzed these systems within the Hubbard model approach, we propose an extension thereof in \ref{ss2}, aiming at obtaining the stabilization of a current-carrying ground state in all of these rings. A possible origin of our extra inter-electronic interaction term is briefly touched upon in section \ref{D}, and will be more carefully addressed in our future work. Finally, we present in \ref{C} a summary of our findings.

\section{Model and Results}\label{MeR}

In order to study the electric transport properties of the eigenstates (in particular the ground state, since we do not take into account thermal effects) of the $N$-site rings in the absence and presence of the uniform and static external magnetic field $\vec{B}=B\hat{z}$ (in this paper, the rings are placed in the $xy$ plane), we used two distinct models: the Hubbard model and our proposed extension of it. Despite the fact that there already exists in the literature at least one work that tried to use the Hubbard model to describe the ring currents in aromatic molecules \cite{Hirsch}, we chose to include it in this work to make it self-contained and also to evidence the need for an extension of this model.

\subsection{Discrete rings according to the Hubbard model} \label{ss1}

The Hubbard model is the simplest model available for dealing with interacting fermions in a crystal lattice. In the absence of an external magnetic field, the Hubbard Hamiltonian for an $N$-site one dimensional lattice, with periodic boundary conditions and with just a single non-degenerate orbital per site, is given by \cite{Essler}
\begin{equation}
	\hat{H}=-t\sum\limits_{j=1}^{N}\sum\limits_{\sigma=\uparrow,\downarrow}\left ( c_{j\sigma}^{\dag}c_{(j+1)\sigma}^{\null}+h.c. \right )+U\sum\limits_{j=1}^{N}\hat{n}_{j\uparrow}\hat{n}_{j\downarrow} \text{ ,}
	\label{ss1eq1}
\end{equation}
 
\noindent where $c_{j\sigma}^{\dag}$ and $c_{j\sigma}$ are the operators that, respectively, create and annihilate an electron with spin $\sigma$ ($\sigma=\uparrow$ for spin up and $\sigma=\downarrow$ for spin down) at the $j$-th site of the ring. Besides, $t$ is the hopping parameter and $U$ is the on-site electronic repulsion. For now, we have not taken into account the next-neighbor electronic repulsion, but we will return to this point later in the text. 

We performed an exact diagonalization of the Hamiltonian Eq.\,(\ref{ss1eq1}) and obtained its eigenvalues and eigenstates $\left \{ \left. \left | \psi_{n}^{(k)} \right. \right \rangle, k=1,2,\cdots,g_{n} \right \}$, with respective degree of degeneracy $g_{n}\geq 1$, for each set of values of $t$, $U$, $3\leq N\leq 6$ and $N_{e}\leq 2N$ initially chosen. It is worth noting that \unexpanded{$\left.\left | \psi_{n}^{(k)} \right. \right \rangle$} are many-body states. Moreover, the quantum number $n$ is a natural number: $n=0$ refers to the ground state and $n\geq 1$, to the $n$-th excited state of the system. Once we have computed the eigenstates of the system, we establish the following criterion: If the matrix representation of the electric current operator (which can be easily found through the continuity equation \cite{Essler}),
\begin{equation}
	\hat{\mathscr{J}}_{el}=-\frac{iet}{N}\sum\limits_{j=1}^{N}\sum\limits_{\sigma}\left ( c_{j\sigma}^{\dag}c_{(j+1)\sigma}^{\null}-c_{(j+1)\sigma}^{\dag}c_{j\sigma}^{\null} \right ) \text{ ,}
	\label{ss1eq2}
\end{equation}

\noindent in the subspace spanned by the $g_{n}$ degenerated eigenstates of the $n$-th excited state of the system (or ground state if $n=0$) is identically zero, then any linear combination of the $g_{n}$ eigenstates, 
\begin{equation}
	\left.\left | \psi \right. \right \rangle=\sum\limits_{k=1}^{g_n}c_{k}\left. \left |\psi_{n}^{(k)}\right.\right\rangle \text{ ,}
	\label{ss1eq3}
\end{equation}

\noindent will lead to a many-body electronic state such that $\left \langle \psi\left | \hat{\mathscr{J}}_{el} \right | \psi\right \rangle=0$. In this situation, the $n$-th excited state is not a current-carrying state in the absence of external fields. On the other hand, if the matrix representation is non-zero, there will be at least one linear combination Eq.(\ref{ss1eq3}) for which $\left \langle \psi\left | \hat{\mathscr{J}}_{el} \right | \psi\right \rangle\neq 0$ and, in this case, we say that the $n$-th excited state can be a current-carrying state in the absence of external magnetic fields. In Eq.\,(\ref{ss1eq2}), $e<0$ is the electronic charge.

Through the preceding analysis, we found that among the excited states of all the rings we studied, there are always some that do not transport electric current and others that, depending on the linear combination we choose in Eq.\,(\ref{ss1eq3}), can be  current-carrying states in the absence of an external magnetic field. For example, in the case of a ring with three sites and two electrons, just the first, second and fourth excited states can be current-carrying states, independently of the value of $U$ we choose. Regarding the ground state of the rings, which is our major interest in this work, we concluded that it can only be a current-carrying state when the number of electrons ($N_e$) of the system is odd, independently of the number of sites of the ring or the parameters $t$ and $U$.  

Since we have determined that there are eigenstates of the rings that can indeed transport electric current, our next step is quantify these currents. To do that, it is necessary to break the degeneracy of the eigenstates, so we could determine, unambiguously, the mean value of the current operator. In order to promote this breakdown of degeneracy, we apply a uniform external magnetic field $\vec{B}=B\hat{z}$ to the ring. The Hamiltonian Eq.(\ref{ss1eq1}) must be modified to incorporate the effects of this field on the electrons of the ring. Because of the minimal coupling $\vec{p}_{j}\rightarrow\vec{p}_{j}-\frac{e}{c}\vec{A}(\vec{r})$, where $e<0$ is the electronic charge, $c$ the speed of light and $\vec{A}(\vec{r})$ the vector potential associated with the magnetic field, each electron in the ring acquires a finite value for its orbital angular momentum, which will originate a persistent current in the sense mentioned in our introductory section. This fact reflects in the Hamiltonian as a gauge transformation of the creation and annihilation operators \cite{Essler}. The other effect is the Zeeman effect, i.e. the coupling between the magnetic field and total spin of the system, which is responsible for the complete breakdown of the degeneracy of the ground states of all the rings we have studied. Therefore, the Hubbard Hamiltonian in the presence of the external magnetic field $\vec{B}=B\hat{z}$ is given by
\begin{equation}
	\begin{matrix}
\hat{H}=-t\sum\limits_{j=1}^{N}\sum\limits_{\sigma}\left (e^{i2\pi f/N}c_{j\sigma}^{\dag}c_{ (j+1)\sigma}^{\null}+h.c. \right )+
\\
U\sum\limits_{j=1}^{N}\hat{n}_{j\uparrow}\hat{n}_{j\downarrow}-\mu_BB\sum\limits_{j=1}^{N}(\hat{n}_{j\uparrow}-\hat{n}_{j\downarrow})
\end{matrix}\text{ ,}
\label{ss1eq4}
\end{equation}
\noindent where the last term refers to the Zeeman coupling and $f=\phi/\phi_{0}$ is the dimensionless magnetic flux that pierces the ring, with $\phi_{0}=hc/\left|e\right|$ being the flux quantum. 

In the presence of the magnetic field, the persistent current that is established in the ground state is given by
\begin{equation}
I(f)=-\frac{c}{\phi_0}\frac{\partial E_0}{\partial f} \text{ .}
\label{ss1eq5}
\end{equation}

\noindent where $E_{0}$ is the ground state energy of Eq.\,(\ref{ss1eq4}) in the absence of the Zeeman coupling (\textit{i.e}, in the Aharonov-Bohm configuration of the field). This is so because the Zeeman term would give a contribution to the current due to the spin of the electrons, whereas we are interested only in the current originated by the orbital movement of the electrons. As a consequence of the Feynmann-Hellman theorem \cite{Sticlet}, the persistent current can also be evaluated through the expectation value  of the current operator in the presence of the field,
\begin{equation}
	\hat{\mathscr{J}}_{el}^{(mag)}=\!-\frac{iet}{N}\!\sum\limits_{j=1}^{N}\!\sum\limits_{\sigma}\!\left ( e^{i2\pi f/N}c_{j\sigma}^{\dag}c_{(j+1)\sigma}^{\null}\!\!-h.c. \right )  \text{ ,}
	\label{ss1eq5p2}
\end{equation}

\noindent in the ground state of the Hamiltonian Eq.(\ref{ss1eq4}). The expression Eq.(\ref{ss1eq5p2}) differs from Eq.(\ref{ss1eq2}) by the complex phase $e^{i2\pi f/N}$, which express the minimal coupling, as discussed above.

Our results revealed that the persistent current is periodic with the magnetic flux $\phi$, with the periodicity being $\phi_{0}/2$ for rings with an odd number of sites at half-filling, and $\phi_{0}$ otherwise, a fact already known in the literature \cite{Maiti,Bobo}. It is evidenced by the blue (solid) curves in the figure Fig.\ref{figI} for rings with (a) three sites and two electrons, (c) three sites and three electrons and (e) six sites and six electrons. We observed that $I(f)$ tends to a finite value as $f$ approaches zero only when the number of electrons in the ring is odd, independently of the number of sites, corroborating our previous results in the absence of the magnetic field. We also studied the behavior of the persistent current as a function of the interaction parameter $U$ and observed, as shown by the blue (solid) curves of the Fig.\,\ref{figI}(b), (d) and (f), that it decreases, as expected, with increasing $U$, since the larger the repulsion $U$, the greater is the tendency of the electron to be localized around the sites.  However, the magnitude of the persistent current is very small in all the microscopic rings we studied (see the Supplementary Material). For example, in the case of the prototype of the benzene molecule, even in a field of $2T$, the persistent current was of the order of $10^{-7} \,eV$ (in natural units, $\hbar=c=1$), too small to account for the anisotropy of the magnetic susceptibility found for this aromatic molecule, as already pointed out in Heirch \textit{et. al.} \cite{Hirsch}. 

Indeed, evaluating the magnetic susceptibility (per mole) due to the current loop Eq.(\ref{ss1eq5}) that is established in the ring, 
\begin{equation}
	\chi_{mag}^{(mol)}=N_{A}\chi=N_{A}\frac{N^4a^4e}{32\pi^{3}\hbar c^{2}}\frac{\partial I(f)}{\partial f} \text{ ,}
	\label{ss1eq6}
\end{equation}
\noindent which is already a measure of the anisotropy of the magnetic response of our rings (since Eq.(\ref{ss1eq6}) takes into account only the orbital degree of freedom of the electrons and the spin contribution to the magnetic susceptibility is isotropic in space in the particular case of our systems), we found for the prototype of the benzene molecule with the realistic values of parameters \cite{Schuler} $t=2,5\,eV$, $U=10\,eV$, and with lattice spacing\cite{Pauling} $a=1,4$\AA, the value $\chi_{mag}^{(mol)}=-1,9\times 10^{-5} cm^{3}/mol$. This value is about three times smaller than the experimental value\cite{And}, $\Delta\chi=-6,49\times 10^{-5}cm^{3}/mol$.  Even when we take into account the inter-atomic next-neighbor repulsion between electrons by adding to the Hamiltonian Eq.(\ref{ss1eq1}) the term $\frac{V}{2}\sum\limits_{<i,j>}\hat{n}_{i}\hat{n}_{j}$ (where $V$ is the next-neighbor interaction parameter, $<i,j>$ represents the next-neighbor sites and $\hat{n}_{j}=\hat{n}_{j\uparrow}+\hat{n}_{j\downarrow}$) and repeat all the previous steps, we find, for\cite{Parr} $V=7,2\,eV$, $\chi_{mag}^{(mol)}=-1,6\times 10^{-5}cm^{3}/mol$, again very different from the experimental value. 

It is not a surprising fact that the Hubbard model cannot account for the anisotropy of the magnetic susceptibility of the aromatic molecules: this model was originally developed to describe the narrow $d$ and $f$ energy bands\cite{Hubbard}, while, in the case of benzene, we have broader energy splittings and, consequently, it is not expected that this model be suitable to describe certain properties of this molecule. In particular its magnetic anisotropy. It is necessary go beyond the Hubbard model to build a microscopic model for the ring currents in aromatic molecules.

\subsection{Discrete rings according to an extension of the Hubbard model}\label{ss2}

\begin{figure}[t!]
\centering
\includegraphics[scale=0.7]{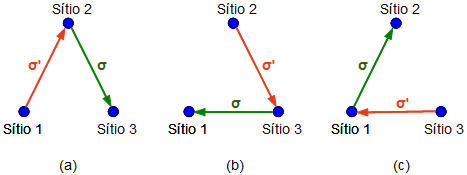}
\caption{\footnotesize{Illustration of the action of the operators (a) $c_{3\sigma}^{\dag}c_{2\sigma'}^{\dag}c_{1\sigma'}^{\null}c_{2\sigma}^{\null}$, (b) $c_{1\sigma}^{\dag}c_{3\sigma'}^{\dag}c_{2\sigma'}^{\null}c_{3\sigma}^{\null}$ and (c) $c_{2\sigma}^{\dag}c_{1\sigma'}^{\dag}c_{3\sigma'}^{\null}c_{1\sigma}^{\null}$ on the electrons in a ring with three sites. The Hermitian conjugation of these operators are responsible for the motion of the electrons in the opposite direction and, for this reason, they are not represented in this figure.}}
\label{ss2fig1}
\end{figure} 

We propose an extension of the Hubbard model, with an \textit{ad hoc} extra inter-electronic interaction term, 
\begin{equation}
 \hat{H}_{I}=-\sum\limits_{\sigma,\sigma'}\Delta_{\sigma\sigma'}\sum\limits_{j=1}^{N}\left ( c_{j\sigma}^{\dag}c_{(j-1)\sigma'}^{\dag}c_{(j-2)\sigma'}^{\null}c_{(j-1)\sigma}^{\null}+h.c. \right ) \text{ ,}
\label{ss2eq1}
\end{equation}

\noindent added to the Hubbard Hamiltonian Eq.(\ref{ss1eq1}). In Eq.(\ref{ss2eq1}), $\sigma$ and $\sigma'$ refer to the electronic spin and $\Delta_{\sigma,\sigma'}$ is a positive adjustable parameter with dimensions of energy,
\begin{equation}
\Delta_{\sigma\sigma'}=\left\{\begin{matrix}
\Delta_1\text{ ,} & \text{ if } \sigma=\sigma'  \\ 
\Delta_2\text{ ,} & \text{ if } \sigma\neq\sigma' 
\end{matrix}\right. \text{ .}
\label{ss2eq2}
\end{equation}

\noindent We built this term with the intention to privilege, energetically, the transport of electric current in the rings. By construction, the interaction $\hat{H}_{I}$ destroys an electron at the $j$-th site of the ring and creates it at some of the next-neighbor sites, promoting an ordered motion of the electrons of the system, which characterizes an electric current. Figure \ref{ss2fig1} illustrates how $\hat{H}_{I}$ acts on the electrons in a ring with three sites. Since Eq.(\ref{ss2eq1}) has a global minus sign, we expected that this term could lower the energy of current-carrying states and, depending on the choice of the parameters $\Delta_1$ and $\Delta_2$, lead to a current-carrying ground state in all of the rings in the absence of an external magnetic field.

\begin{figure}[t!]
\centering
\includegraphics[scale=0.75]{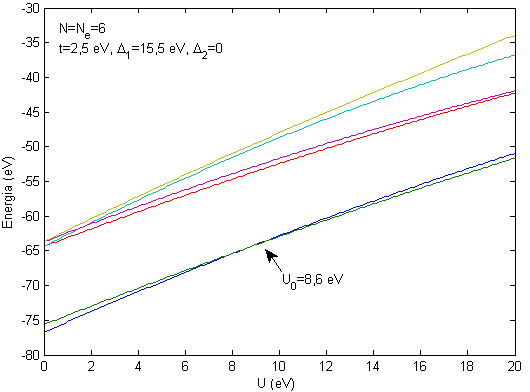}
\caption{\footnotesize{Fist six levels of the energy spectrum of the Hubbard Hamiltonian added to $\hat{H}_{I}$, as a function of $U$, for the prototype of the benzene molecule ($N=N_e=6$). We used $t=2,5 \,eV$, $\Delta_1=15,5\,eV$ and $\Delta_2=0$.}}
\label{ss2fig2}
\end{figure} 

The study of the electric transport properties of the rings in this case follows step-by-step what was done in subsection \ref{ss1}. We repeat the numerical approach and, hereafter, our attention lies in the ground state of the rings. We performed an exact diagonalization of the Hubbard Hamiltonian with the extra interaction term Eq.\,(\ref{ss2eq1}) and, through the comparison of its energy spectrum as a function of $U$ with the spectrum of Eq.\,(\ref{ss1eq1}), we found that, for all of the rings studied, $\hat{H}_{I}$ promotes a degeneracy breakdown of some of its eigenstates.

Furthermore, we observed that it is always possible to choose $\Delta_1$ and $\Delta_2$ for which there is a energy level crossing between the ground state and one of the excited states of the system. $U_{0}$ denotes the value of the on-site repulsion parameter where the crossing takes place. In Fig.\,\ref{ss2fig2} we can see, for example, the level crossing between the ground state and first excited state of the prototype of the benzene molecule ($N=N_e=6$), with $\Delta_1=15,5\,eV$ and $\Delta_2=0$. Crossings involving the ground state level cannot be found in the Hubbard Hamiltonian in the absence of $\hat{H}_{I}$, and the values of $\Delta_1$ and $\Delta_2$ for which they happen are not unique: for example, in a ring with four sites and three electrons a level crossing between the ground state and first excited state occurs either for $(\Delta_1;\Delta_2)=(0,5;0,5)\,eV$, with $U_0=6,5\, eV$, or $(\Delta_1,\Delta_2)=(0,1)\,eV$ with $U_{0}=7,2\,eV$. 

\begin{figure}[t!]

\centering
\includegraphics[scale=0.4]{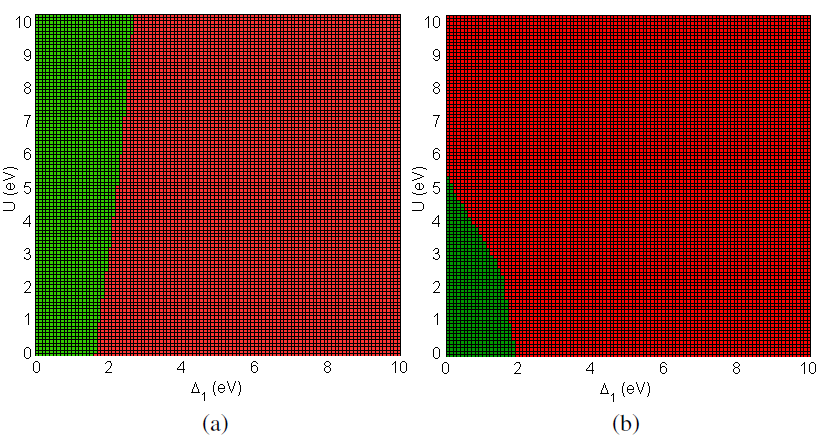}
\caption{\footnotesize{Color maps of the space parameter $\Delta_1\times U$ for a ring with (a) $N=3$ sites and $N_e=3$ electrons with $t=1\,eV$ and $\Delta_2=0,8 \,eV$ fixed and (b) $N=4$ sites and $N_e=3$ electrons with $t=1\,eV$ and $\Delta_2=1,5\,eV$ fixed. The red regions represent points $(\Delta_1,U)$ of the space parameter for which the matrix representation ($M$) of the current operator, Eq.\,(\ref{ss2eq3}), in the subspace of the ground state is zero. Whereas the green  regions are those points for which $M\neq 0$.}}
\label{ss2fig3} 
\end{figure}

A question that naturally arises is: is the new ground state after the crossing (\textit{i.e.} the ground state of the ring for $U>U_{0}$) a current-carrying state? To answer this question, we have to analyze the matrix $M$ representing the current operator in the subspace spanned by the $g_{0}\geq 1$ eigenstates relative to the ground state of the ring before and after the crossing, using the same criterion presented in the former subsection: if the matrix representation is identically zero, the ground state is not a current-carrying state. Otherwise, the ground state can support an electric current, depending on the linear combination of the $g_{0}$ eigenstates we choose. However, Eq.\,(\ref{ss1eq2}) is no longer the correct form of the current operator. This is because $\hat{H}_{I}$ does not commute with the number operator, $\hat{n}_{j\sigma}=c_{j\sigma}^{\dag}c_{j\sigma}^{\null}$, and, therefore, in order to obey the continuity equation, the expression for the current operator must be modified:
\begin{equation}
	\begin{matrix}
\hat{\mathscr{J}}_{el}=-\frac{iet}{N}\sum\limits_{j=1}^{N}\sum\limits_{\sigma}\left ( c_{j\sigma}^{\dag}c_{(j+1)\sigma}^{\null}-h.c. \right )+\\ 
+\frac{2ie}{N}\sum\limits_{\sigma,\sigma'}\Delta_{\sigma\sigma'}\sum\limits_{j=1}^{N}\left ( c_{j\sigma}^{\dag}c_{(j-1)\sigma'}^{\dag}c_{(j-2)\sigma'}^{\null}c_{(j-1)\sigma}^{\null} -h.c.\right ) \text{ .}
\end{matrix}
\label{ss2eq3}
\end{equation}

\begin{figure*}[t!]
\centering
\includegraphics[scale=0.7]{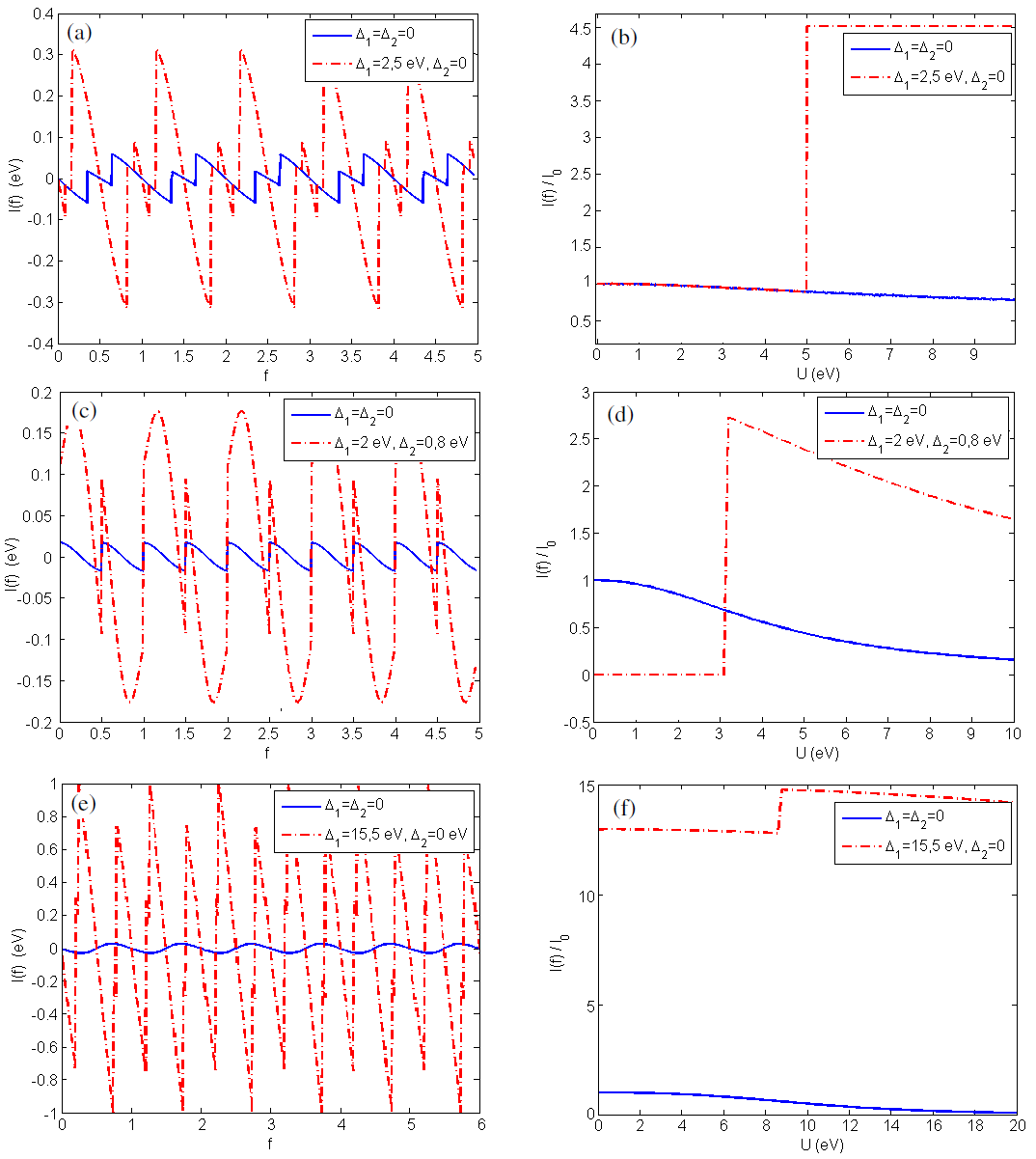}
\caption{\footnotesize{Persistent current, as a function of the magnetic flux that pierces a ring of (a) three sites and two electrons with $t=1\,eV$ and $U=8\,eV$, (c) three sites and three electrons with $t=1\,eV$ and $U=6\,eV$ and (e) six sites and six electrons (prototype of the benzene molecule) $t=2,5\,eV$ and $U=10\,eV$. In figures (b), (d) and (f) it is shown the persistent current as function of the on-site repulsion for the same system of the figures (a), (c) and (e), respectively and with $B=2\,T$ fixed. The blue (see the colored version of the paper) continuous lines refer to the Hubbard model without the extra interaction term $\hat{H}_{I}$ (\textit{i.e.} $\Delta_{1}=\Delta_2=0$), whereas the dashed red curves refers to $\Delta_{\sigma,\sigma'}\neq 0$. The values of $\Delta_{1}$ and $\Delta_2$ chosen in each case are indicated in the figures. In order to ease the comparison between the persistent currents obtained with the Hubbard model and our extension thereof, both of them are normalized by $I_{0}$, the value of the persistent current in the absence of any electronic interaction (pure hopping model). We set the lattice spacing $a=1$\AA \,in all figures, except in (e) and (f), where $a=1,4$\AA.}}
\label{figI}
\end{figure*}

\noindent This expression can be easily derived and the demonstration can be found in the Supplementary Material. Since our model has two adjustable parameters, $\Delta_1$ and $\Delta_2$ (besides $t$ and $U$) a careful analysis of the parameter space had to be done in order to find whether the ground state could be a current-carrying state: we chose several values of $\Delta_1$ and $U$ (with $N$, $N_{e}$, $t$ and $\Delta_2$ fixed) and for each pair $(\Delta_1,U)$, we performed the diagonalization of the total Hamiltonian of the system and evaluated the matrix representation (denoted here by $M$) of Eq.\,(\ref{ss2eq3}) in the subspace corresponding to the ground state of the total Hamiltonian. Thus, we built a color map of the parameter space $\Delta_{1}\times U$, where the green regions represent points for which $M\neq 0$ and the red regions those for which $M=0$. In Fig.\ref{ss2fig3}, we have two of these maps, for rings with (a) three sites and three electrons and (c) four sites and three electrons. Other examples of these maps can be found in the Supplementary Material. Contrary to our expectation, the extra interaction Eq.\,(\ref{ss2eq1}) was not able to stabilize a current-carrying ground state in any of the rings we studied. Although we could always promote a level crossing between the ground state and some excited state of the rings, the new ground state for $U>U_{0}$ was not necessarily a current-carrying state. We observed different kinds of behavior, depending on the number of the electrons in the rings: similarly to subsection \ref{ss1}, for rings with an even number of electrons we could never stabilize a current-carrying ground state in the absence of an external field, independently of the number of sites of the ring and values of the parameters $t$, $U$, $\Delta_1$ and $\Delta_2$ chosen. On the other hand, for rings with $N_{e}$ odd and at half-filling or above ($N_{e}\geq N$) the behavior was in agreement with what we initially expected for all the rings: beyond the level crossing, the ground state starts to carry an electric current, \textit{i.e.}, in this case the extra interaction term $\hat{H}_{I}$ has successfully stabilized a current-carrying ground state. Alternatively, for rings with $N_{e}$ odd and below half-filling ($N_e<N$) the behavior was opposed to the former: the ground state could only be a current-carrying state before the level crossing. These results suggest an strong dependence on the number of electrons of the system. The dependence of the properties of microscopic Hubbard rings with the number of electrons was already reported in another work\cite{Maiti}.
 
\begin{figure*}[t!]
\begin{minipage}{0.5\textwidth}
\centering
\includegraphics[scale=0.55]{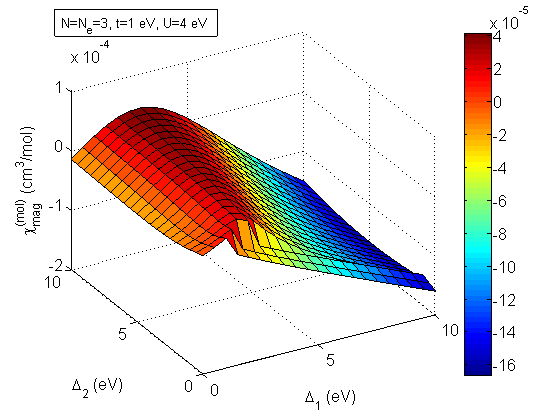}
\begin{center}
\footnotesize{(a)}
\end{center}
\end{minipage}%
\begin{minipage}{0.5\textwidth}
\centering
\includegraphics[scale=0.55]{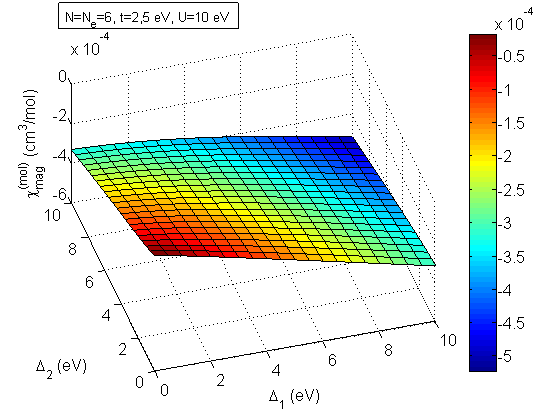}
\begin{center}
\footnotesize{(b)}
\end{center}
\end{minipage}
\caption{\footnotesize{Molar magnetic susceptibility, Eq.\,(\ref{ss1eq6}) as function of the parameters $\Delta_1$ and $\Delta_2$ for (a) a ring with $N=3$ sites and $N_{e}=3$ electrons, $t=1\,eV$ and $U=4\,eV$ and (b) the prototype of the benzene molecule ($N=N_{e}=6$) with $t=2,5\,eV$ and $U=10\,eV$. The value of the lattice spacing parameter we used was (a) $a=1$\AA \,and (b) $a=1,4$\AA.}}
\label{ss2fig4}
\end{figure*}

Nevertheless, the most relevant result of our work was found in the presence of the external magnetic field $\vec{B}=B\hat{z}$. In this case, the extra interaction term also acquires a complex phase because of the gauge transformation of the electronic creation and annihilation operators \cite{Essler} and the total Hamiltonian of the system is given by
\begin{align}
\begin{split}
\hat{H}= -t\!\sum\limits_{j=1}^{N}\sum\limits_{\sigma}\left (e^{i\theta_{1}\!(f)}c_{j\sigma}^{\dag}c_{ (j+1)\sigma}^{\null}+h.c. \right )+ U\!\sum\limits_{j=1}^{N}\hat{n}_{j\uparrow}\hat{n}_{j\downarrow}+\\ 
-\!\!\sum\limits_{\sigma,\sigma'}\!\Delta_{\sigma\sigma'}\!\!\sum\limits_{j=1}^{N}\!\left (e^{-i\theta_{2}\!(f)} c_{j\sigma}^{\dag}c_{(j-1)\sigma'}^{\dag}c_{(j-2)\sigma'}^{\null}c_{(j-1)\sigma}^{\null} + h.c.\right )+ \\ 
-2\mu_B B\hat{S}_{z}\;\;\;\;\;\;\;\;\;\;\;\;\;\;\;\;\;\;\;\;\;\;\;\;\;\;\;\;\;\;\;\;\;\;\;\;\;\;\;\;\;\;\;\;\;\;\;\;\;\;\;\;\;\;\;\;\;\;\;\;\;\;\;\;\text{ ,}
\label{ss2eq4}
\end{split}
\end{align}


\noindent where $\theta_{1}(f)=2\pi f/N$ and $\theta_{2}(f)=4\pi f/N$. Similarly to what we discussed in subsection \ref{ss1}, the Zeeman coupling term in Eq.\,(\ref{ss2eq4}) is responsible for the complete breakdown of the degeneracy of the ground state of the rings. Moreover, the persistent current in their ground states, Eq.(\ref{ss1eq5}) (where $E_{0}$ is now the energy of the ground state of the Hamiltonian Eq.\,(\ref{ss2eq4}) without the Zeeman term), is also periodic with the flux that pierces the ring. In this case, however, contrary to the results of the subsection \ref{ss1}, the periodicity is always $\phi_{0}$, independently of the number of sites and electrons of the ring. See, for example the differences in the periodicity of the blue (solid, with $\Delta_1=\Delta_2=0$) and red (dashed, with $\Delta_1=2\,eV$ and $\Delta_2=0,8\,eV$) curves of the Fig.\,(\ref{figI})(c). Besides, when we compare the persistent current as a function of $U$ obtained through our proposed model with those evaluated in subsection \ref{ss1}, we find that the extra interaction term enhances the magnitude of the persistent current in all the rings studied, including those with an even number of electrons, for which the ground state was not a current-carrying state in the absence of the external magnetic field. The enhancement of the persistent current is evidenced in Fig.\,\ref{figI}(b), (d) and (f): we can see that beyond the level crossing between the ground state and some excited state of the system (\textit{i.e.} for $U>U_{0}$) the persistent current is considerably more intense than that before the crossing. In addition to the enhancement of the magnitude of the persistent current, our extra inter-electronic interaction term $\hat{H}_{I}$ appears to amplify the diamagnetic response of the rings. In particular, for the prototype of the benzene molecule, if we set $\Delta_1=1.5\,eV$ and $\Delta_2=0$, with the realistic values $t=2,5\,eV$ and $U=10\,eV$ for the hopping and the on-site repulsion parameters, we can recover the order of magnitude of the magnetic anisotropy observed experimentally for this molecule, obtaining, through Eq.(\ref{ss1eq6}), $\chi_{mag}^{(mol)}=-5,9\times 10^{-5}cm^{3}/mol$. The graph of $\chi_{mag}^{(mol)}$ as function of both $\Delta_1$ and $\Delta_2$ in Fig.\ref{ss2fig4}\,(b) evidences the amplification of the diamagnetic response for the prototype of the benzene molecule: increasing $\Delta_1$ or $\Delta_2$ increases the magnitude of $\chi_{mag}^{(mol)}$, which is negative in this particular case.  

The amplification of the diamagnetic response was also observed for rings other than the prototype of the benzene molecule. However, in some cases (for example the ring with three sites and three electrons), we find what appears to be a competition between a paramagnetic and a diamagnetic response. As shown in Fig.\,\ref{ss2fig4}(a), if we fix $\Delta_1=2\,eV$ and increase $\Delta_{2}$, $\chi_{mag}^{(mol)}$ also increases, becoming positive (which characterizes a paramagnetic response) for large values of $\Delta_2$. 

\section{Discussion}\label{D}

In the previous section, we showed through a numerical approach (an analytic procedure is impracticable, since the dimension of the state space grows with the factorial of the number of sites and electrons of the rings) that adding the extra interaction term $\hat{H}_{I}$ in Eq.\,(\ref{ss2eq1}) to the Hubbard Hamiltonian Eq.\,(\ref{ss1eq1}) leads to more intense persistent currents in the ground state of the rings when they are subject to an external static magnetic field $\vec{B}=B\hat{z}$ perpendicular to their planes. Moreover, $\hat{H}_{I}$ is able to amplify the diamagnetic response of the rings. In particular, for the ring with six sites and six electrons, we recovered the order of magnitude of the magnetic anisotropy observed experimentally for the benzene molecule. As mentioned in subsection \ref{ss2}, Eq.\,(\ref{ss2eq1}) is an \textit{ad hoc} interaction term and a question that remains is: what could be the microscopic origin of $\hat{H}_{I}$? In this section we will briefly touch upon this subject.

The first possibility is that $\hat{H}_{I}$ could come from the Coulomb repulsion itself, but we will show that this is not the case. The many-body Hamiltonian for a system of $N_{e}$ electrons on a lattice is
\begin{equation}
	H=\sum\limits_{i=1}^{N_e}\left ( \frac{\vec{P}_{i}^{2}}{2m}+V(\vec{r}_{i}) \right )+\frac{1}{2}\mathop{\sum\limits_{i,j=1}^{N_e}}_{j\neq i}\frac{e^2}{\left|\vec{r}_{i}-\vec{r}_{j}\right|}\text{ ,}
\label{Deq1}
\end{equation}

\noindent where $V(\vec{r}_{1})$ is the periodic potential of the lattice. In the formalism of second quantization, Eq.\,(\ref{Deq1}) becomes
\begin{equation}
 \hat{H}=\sum\limits_{i,j=1}^{N}\sum\limits_{\sigma}t_{ij}\,c_{ i,\sigma}^{\dag}c_{j,\sigma}^{\null}+\frac{1}{2}\sum\limits_{i,j,k,l}\sum\limits_{\sigma,\sigma'}U_{ijkl}\,c_{i,\sigma}^{\dag}c_{j,\sigma'}^{\dag}c_{k,\sigma'}^{\null}c_{l,\sigma}^{\null}	\label{Deq2}\text{ ,}
\end{equation}

\noindent with
\begin{equation}
t_{ij}=\int d^{3}r \, \phi_{i}^{*}(\vec{r})\left [ \frac{P^{2}}{2m}+V(\vec{r}) \right ]\phi_{j}(\vec{r}) \text{ ,}
	\label{Deq3}
\end{equation}

\noindent being hopping integrals, where $\phi_{j}(\vec{r})=\phi(\vec{r}-\vec{R}_{j})$ are the Wannier wave function localized around the $j$-th site of the lattice (with position vector $\vec{R}_{j}$) and 
\begin{equation}
U_{ijkl}=\int\!\!\!\int d^{3}r\,d^{3}r'\phi_{i}^{*}(\vec{r})\phi_{j}^{*}(\vec{r}\,')\frac{e^2}{\left | \vec{r}-\vec{r}\,' \right |}\phi_{k}(\vec{r}\,')\phi_{l}(\vec{r}) \text{ ,}
	\label{Deq4}
\end{equation}

\noindent the inter-electronic interaction integrals. The Hubbard Hamiltonian Eq.(\ref{ss1eq1}) is obtained by making some approximations on Eq.\,(\ref{Deq3}) and Eq.\,(\ref{Deq4}): firstly, we assume that $t_{ij}=-t$ if $i$ and $j$ are nearest neighbors sites and $t_{ij}=0$ otherwise. The second approximation consists in neglecting all the interaction integrals other than the leading one $U_{iiii}\equiv U$, which is nothing but the on-site repulsion. However, if we take into account, besides $U_{iiii}$, interactions integrals with $U_{i,i-1,i-2,i-1}$ and $U_{i-1,i-2,i-1,i}=U_{i,i-1,i-2,i-1}^{*}$, we find, if we assume that $U_{i-1,i-2,i-1,i}\equiv \Delta$ has the same value for all sites of the ring and also is a real parameter, the following interaction term 
\begin{equation}
	\hat{H}_{2}=\frac{1}{2}\Delta\sum\limits_{j=1}^{N}\sum\limits_{\sigma,\sigma'}\left ( c_{j\sigma}^{\dag}c_{(j-1)\sigma'}^{\dag}c_{(j-2)\sigma'}c_{(j-1)\sigma} +h.c.\right ) \text{ .}
	\label{Deq6}
\end{equation}

\noindent Comparing Eq.\,(\ref{ss2eq1}) and Eq.\,(\ref{Deq6}), we find that our extra interaction term $\hat{H}_{I}$ coincides with $\hat{H}_{2}$ only if $\Delta_1=\Delta_2=\frac{\Delta}{2}$ and $\Delta<0$. Besides, the overlap integral $U_{i,i-1,i-2,i-1}$ is generally much smaller than the leading term $U_{iiii}$, as argued by Hubbard \textit{et. al.} \cite{Hubbard}, while we used $\Delta_1$ and $\Delta_2$ comparable with $t$ and $U$. Consequently, the extra interaction term Eq.\,(\ref{ss2eq1}) cannot come from the neglected terms of the Coulomb interaction.  

Another attempt to explain the origin of that term comes from the explicit form of the electric current from the $(i-1)$-th to the $i$-th site of the ring which reads 
\begin{equation}
J_{i-1,i}=-iet\sum\limits_{\sigma}(c_{i\sigma}^{\dag}c_{(i-1)\sigma}-h.c.) \text{ .}
	\label{Deq7}
\end{equation}

\noindent Therefore, multiplying $J_{i-1,i}$ and $J_{i-2,i-1}$ we obtain 
\begin{align}
\begin{split}
J_{i-1,i}J_{i-2,i-1}=- e^2t^2\sum\limits_{\sigma} c_{i\sigma}^{\dag}c_{(i-2)\sigma}+
\\ 
 + e^2t^2\sum\limits_{\sigma}\left (  c_{i\sigma}^{\dag}c_{(i-2)(-\sigma)}^{\dag}c_{(i-1)(-\sigma)}c_{(i-1)\sigma}+h.c.   \right ) +\\ 
 - e^2t^2\sum\limits_{\sigma,\sigma'}\left ( c_{i\sigma}^{\dag} c_{(i-1)\sigma'}^{\dag} c_{(i-2)\sigma'} c_{(i-1)\sigma}+h.c.    \right )\text{ .} 
	\label{Deq8}
\end{split}
\end{align}

\noindent The last term of Eq.\,(\ref{Deq8}) has the same structure of Eq.\,(\ref{ss2eq1}). Thus we believe that our extra interaction could be at least one of the terms of a current-current interaction. However, this current-current interaction cannot be the conventional electromagnetic one (the Breit-Darwin interaction\cite{Darwin,BreitI,BreitII}), because the latter is a relativistic correction to the Coulomb interaction and, for this reason, its magnitude is of the order of $v^2/c^2$ of the characteristic molecular energy scales, and, consequently, much smaller than the values of $\Delta_1$ and $\Delta_2$ ($\sim 1\,eV$ to $10\,eV$) we used throughout our work. Instead, inspired by the fact that $\Delta_1$ and $\Delta_2$ we used are comparable with $U$, the current-current-like interaction we are looking for should be mediated by the electronic matter itself, possibly by the bond $\sigma$-electrons, which occupy the hybrid $sp^{2}$ orbitals in the plane of the rings (equivalent to the $\sigma$-orbitals of the carbon rings in aromatic molecules). If the influence of these fast core electrons over the slow $\pi$-electrons of the $p_{z}$ orbitals can be written as a vector potential (similarly to the correction to the adiabatic theorem within the Bohr-Oppenheimer scheme), we could hopefully write an interaction mediated by the bond electrons that would have the same structure of a current-current interaction but much more intense than the relativistic correction. We are planning to explore these ideas elsewhere in the near future.

\vspace{1cm}
\section{Conclusions}\label{C}

The aim of our work was to investigate the electronic transport properties of small discrete rings with $3\leq N\leq 6$ sites and $N_{e}<2N$, both in the presence and absence of an external uniform and static magnetic field perpendicular to the plane of the rings, \text{i.e.} $\vec{B}=B\hat{z}$, in the light of two microscopic model: the Hubbard model and our proposed extension of it. Our results within the Hubbard model confirmed what was already known in the literature: this model canmot account for the anisotropy of aromatic molecules. We obtained, using realistic values for the hoping and on-site repulsion parameters $t=2,5\,eV$ and $U=10\,eV$, as well as a lattice spacing $a=1,4$\AA, a diamagnetic anisotropy of $-6,49\times 10^{-5}cm^{3}/mol$, which is about three times smaller than the experimental value. Regarding our extension of the Hubbard model, which consists of an \textit{ad hoc} extra inter-electronic interaction, $\hat{H}_{I}$ (see Eq.(\ref{ss2eq1})), we found that, contrary to our initial expectation, it could not stabilize a current-carrying ground state in any of the rings studied in the absence on a external magnetic field, although we could always find a level crossing between the ground state and some of the excited states of the system. However, in the presence of the field $\vec{B}=B\hat{z}$, we found that $\hat{H}_{I}$ enhances the persistent current in the ground states of the rings (compared to the persistent currents evaluated with the Hubbard model Eq.\,(\ref{ss1eq1})) and also the component of the diamagnetic susceptibility of the rings parallel to the field, \textit{i.e.}, $\hat{H}_{I}$ appears to amplify the diamagnetic response of the rings. In particular for the prototype of the benzene molecules, when we set $t=2,5\,eV$, $U=10\,eV$, $\Delta_1=1,5\, eV$ and $\Delta_2=0$, we find a magnetic anisotropy of $-5,9\times 10^{-5}cm^{3}/mol$ and thus we recovered the order of magnitude of diamagnetic anisotropy measured experimentally for this molecule. 

Our studies revealed a rich physics, with non-trivial energy spectrum of the rings, level crossings induced by our extra interaction term, strong dependence on the number of electrons of the system and even a possible competition between the diamagnetic and paramagnetic responses depending on the values of the adjustable parameter $\Delta_1$ and $\Delta_2$. Based on our results, we believe that our model, although simplified, can help us to understand some aspects of the physics behind the ring currents in aromatic molecules. An important point to be addressed in the near future is a more rigorous investigation on the possible origin of $\hat{H}_{I}$, searching, in particular, for a current-current-like interaction mediated by the electronic matter itself.    

\section*{Supplementary Material}

See Supplementary Material for more details about the energy spectra of the rings either in the presence or absence of $\hat{H}_{I}$, as well as the behavior of the persistent current that is established in the ground state of the rings as a function of the magnetic flux $f$ and the on-site repulsion. The reader can also find in this material other examples of the maps of the parameter space $\Delta_{1}\times U$ and a demonstration of the expressions for the current operator Eq.\,(\ref{ss1eq2}) and Eq.\,(\ref{ss2eq3}).

\section*{Acknowledgments}

We acknowledge the Conselho Nacional de Desenvolvimento Cient\'ifico e Tecnol\'ogico (CNPq) for the financial support.

\bibliography{RefsArticleII}{}
\bibliographystyle{plain}


\end{document}